# From Cues to Signals: Evolution of Interspecific Communication Via Aposematism and Mimicry in a Predator-Prey System

Kenna D. S. Lehmann[1,3,4,*], Brian W. Goldman[2,3], Ian Dworkin[1,3,4], David M. Bryson[3], Aaron P. Wagner[3]

1 Department of Zoology, Michigan State University, East Lansing, MI, USA

2 Department of Computer Science and Engineering, Michigan State University, East Lansing, MI, USA

3 BEACON Center for the Study of Evolution in Action, Michigan State University, East Lansing, MI, USA

4 Program in Ecology, Evolutionary Biology, and Behavior, Michigan State University, East Lansing, MI, USA

* E-mail: Corresponding kdslehmann@gmail.com

## Abstract

Current evolutionary theory suggests that many natural signaling systems evolved from preexisting cues. In aposematic systems, prey warning signals benefit both predator and prey. When the signal is highly beneficial, a third species often evolves to mimic the unpalatable species, exploiting the signaling system for its own protection. We investigated the evolutionary development of predator cue utilization and prey signaling in a digital predator-prey system in which mimicking prey could evolve to alter their appearance to resemble poison-free or poisonous prey. In predators, we observed rapid evolution of cue recognition (i.e. active behavioral responses) when presented with sufficiently poisonous prey. In addition, active signaling (i.e. mimicry) evolved in prey under all conditions that led to cue utilization. Thus we show that despite imperfect and dishonest signaling, given a high cost of consuming poisonous prey, complex systems of interspecific communication can evolve via predator cue recognition and prey signal manipulation. This provides evidence supporting hypotheses that cues may serve as stepping-stones in the evolution of more advanced communication systems and signals incorporating information about the environment.

## Introduction

Signaling systems represent a basic form for inter- and intraspecific communication. Signals are an evolved means of actively conveying information and influencing the behavior of receivers. In contrast, cues are passive, non-evolving biological and environmental traits that inherently provide the observer with information. Organisms have evolved to use both signals and cues to inform their behavior. One of the most pervasive examples of signaling systems in the animal world is aposematism: the warning coloration of poisonous and distasteful species. Aposematism occurs in a wide variety of taxa [1–3] and has been heralded as one of the most striking examples of evolution [4]. The characteristics of these aposematic signaling systems are highly variable. Chief among these, the toxicity of the aposematic prey (i.e. the model) and the occurrence, palatability, and accuracy of mimics are disparate across systems. Studies of a variety of aposematic systems have provided insight into a vast number of evolutionary concepts including character displacement [5], frequency dependence [6,7], diversity [8,9], gene flow [10], co-evolution and co-evolutionary arms races [11,12], and adaptive landscapes [13]. However, despite these studies and the development of theoretical models on aposematic signaling and mimicry[14–21], we do not fully understand the conditions necessary for their evolution because no studies have been able to investigate their evolution from a primordial system. To complicate matters further, predator learning is not fully understood and its importance for the evolution of aposematism and mimicry likely varies between species.

Current theory suggests that evolution of the aposematic signaling in unpalatable species is highly adaptive because it benefits both predator and prey [15]: prey benefit when predators learn to avoid them, predators benefit by avoiding harmful prey. Such signaling systems often include one or more species that mimic the aposematic signal in order to avoid predator attacks. In Müllerian mimicry, a number of species share a common warning signal to advertise unpalatability. The species within such mimicry rings share the costs associated with educating a common predator [22]. In contrast, Batesian mimics advertise a warning signal while remaining palatable [23]. This dishonest signal benefits only the mimic and is detrimental to the unpalatable signaler and the receiving predator because it degrades the information quality of the aposematic signal.

In terms of accuracy of mimic signals, for both Müllerian and Batesian mimicry, predation on imperfect mimics is expected to select for perfect mimicry, or automimicry (e.g., [24]). For automimicry systems, quantitative models suggest that levels of protection enjoyed by the mimic scale with the noxiousness and palatability of the model prey [25,26]. In addition, a number of experimental studies have shown that accurate mimicry evolves when the model is scarce relative to the abundance of mimics [6].

Although selection often favors automimicry, stable systems of imperfect mimicry are prevalent in nature and occur under many conditions [18–20]. In order to describe the processes by which they can evolve, systems of imperfect mimicry have been widely studied from both the theoretical [21,24,27–32] and experimental perspectives [18,20,25,33–37]. From these, two primary effects seem to support the evolution of imperfect mimics. First, selective pressure for perfect mimicry relaxes when imperfect mimics are rare relative to the model or when imperfect mimics are unprofitable due to other factors (e.g., size or agility) [18,38]. Second, predators also exert less selective pressure on imperfect mimics when they generalize the unpalatable poisonous prey's characteristics, leading to behavioral avoidance of any species that exhibit these generalized traits [18,39,40].

In order to determine the evolutionary pathways leading from a naïve predator-prey system to a fully functional aposematic signaling system, quantitative models have outlined the theoretical conditions under which aposematic signaling systems stabilize (e.g., [13,26,41–50]). However, such models require substantial simplification of the signaling system, assumptions of unnatural conditions, or reliance on extant signaling systems (bringing us full circle). Additionally, no studies have been able to experimentally examine the conditions necessary for the realized evolution of predator-prey signaling systems. This is a difficult challenge, given that, in order to fully evaluate the conditions necessary for the evolution of aposematic signaling, one must observe its evolution in a naïve system where signaling has not yet evolved. However, all predators available for experiments have preexisting and established signal recognition systems. To resolve these issues, we used the digital evolution research platform Avida [51] to test the conditions leading to the evolution of a mimicry signal from an aposematic cue in a coevolutionary predator-prey system. We tested for levels of toxicity necessary for the evolutionary emergence of (1) recognition of signaling cues by predators and (2) dishonest signaling by prey mimics. Highly toxic model species are predicted to support more numerous and less accurate mimics [18–20,38,39]. Further, the maintenance of a dishonest signal, as in Batesian mimicry, is expected to require accurate mimicry when an abundant signal accompanies low toxicity [25,26]. Accordingly, we also tested the level of mimic accuracy required to support a successful Batesian mimic population while varying the levels of model toxicity. We hypothesized that these two conditions (high toxicity or accurate mimicry) provide the necessary selective pressures for dishonest signaling to arise from an existing cue.

## Materials and Methods

We used the digital evolution software Avida to assess the conditions facilitating the evolution of predators that utilize cues to inform their behavior and prey that actively signal, via mimicry, and influence the feeding habits of cue-receptive and cue-sensitive predators. Avida organisms have a sequence of program instructions that controls their behavior and serves as genetic information inherited by their offspring. Instructions executed on the genome dictate the actions taken by an organism, allowing it to sense and interact with the environment (e.g., obstacles, food, other organisms), process information, or reproduce. The genome replication process is imperfect, allowing for the introduction of mutations into offspring genomes. Differential fitness in the populations occurs as a consequence of mutations producing novel combinations of operations. Combined, these properties of Avida provide the conditions necessary for adaptive evolution via natural selection: replication, inheritance, variation, and differential fitness [52]. Over the course of evolution, digital organisms in Avida often exhibit behaviors similar to biological organisms observed in natural systems [44].

We configured Avida to enable predator-prey interactions [53] using a modified form of Avida's Heads-EX hardware [54]. To allow for the evolution of predators, we included an attack instruction that could mutate into an organism's genome, making it capable of consuming other organisms. In our digital ecosystems, prey species consumed spatially distributed, limited resources across a 251x251 grid-cell environment. Once resources in a cell were consumed, the environmental resource was replenished at a rate of 0.01 resource units per cell per update, to a maximum per-cell level of 1 full unit (1 unit = minimum level consumable by prey). Prey organisms could utilize sensory information and movement instructions to locate and reach edible resources. Predators (once evolved) had the same set of potential genetic instructions as prey. Thus predators could evolve to use the same instructions to locate and consume prey, through which they gained all of their previously collected resources. All organisms were required to consume a total of 10 units of resource before they could reproduce, either directly from the environment (prey) or from consuming prey that had consumed resources (predators). Reproduction was thus limited by resource consumption: the faster an organism gathered food, the sooner it could reproduce. Thus, for predators, advantageous mutations were those that allowed for more rapid targeting and capturing of prey. Likewise, any prey mutations that conferred greater foraging efficiency or predator avoidance skills, would give prey a selective advantage.

Prey organisms were divided into three classes of morphs: non-poisonous, poisonous, and mimic. To control for subpopulation size effects, and because predators could not act as a top-down control on the poison prey class, classes were assigned at birth such that 50% were poisonous, 25% were safe, and the remaining 25% were potential mimics. The designated class was a part of the prey's phenotype visible to other organisms. Each morph class foraged for separate environmental resources.

'Non-poisonous' prey organisms directly transferred their gathered resources to predators when consumed. The 'poison' prey organisms, upon consumption, reduced a predator's gathered resources by a factor (i.e. 'poison level', see below) of what that prey had previously gathered. The 'mimic' prey provided the same resource benefit as 'non-poisonous' prey when consumed by a predator. However, mimic prey were unique in that, if it had mutated into their genome, they could execute an instruction that allowed them to change their visible phenotype to that of a different prey class. This instruction had no effect on organisms not in the mimic class. Displayed and visible classes provided a cue for predators, and an evolutionary opportunity for them to evolve abilities for recognizing and avoiding poisonous prey. If these behaviors evolved, mimics would then be able to further evolve to manipulate that cue, avoiding predation by providing the predators with a false signal.

Predators were classified as such as soon as they made their first kill. The prey class preference of each predator was determined by a specific instruction sequence defining the 'attack organism operation' (see Table 1). The default sequence, constituting a single attack instruction, performed a 'generalist' attack, targeting any prey organism in the cell in front of the predator, regardless of the prey's displayed signal. Three additional attack options consisted of an attack instruction followed by one of eight modifying instructions distributed across three classes such that three specified a safe prey attack, three mapped to mimic prey type attack, and two mapped to a poisonous prey attack. If the victim's displayed class did not match the specified attack type (i.e. predator preference), the attack would fail. As a result, a predator's prey preference was explicitly heritable, though multiple preferences could be expressed if multiple attack instruction sequences existed within the predator's genome. Under most treatments, when a predator attacked a mimic, in order to make a kill, the predator's expressed prey preference had to match the mimic's displayed class (i.e. what it mimicked), not its true class. Select treatments, as noted in the results, further altered the fidelity of the apparent prey phenotype such that predators perceived the true class, instead of the displayed class, with the specified probability.

All experiments were started with the introduction of one prey organism from each class. Each organism's genome was 100 instructions long, and each of these ancestors simply moved randomly through the Avida landscape, attempting to consume sufficient resources to reproduce. Genetic mutation rates applied to offspring genomes were a 0.25 probability of a single instruction substitution, and a 0.05 probability each that a single instruction would be inserted or deleted. Assignment of offspring to prey classes ensured that half of all prey born were safe for predators to eat, and that the cue from the poisonous prey outnumbered any signaling by the mimics. Importantly, this does not mean heeding the cue was always advantageous: the reward for eating a signaling mimic may have outweighed the penalty for eating a poisonous prey. Furthermore, as we controlled only birth ratios, it was possible for the number of mimics in a given population to outnumber the poisonous prey. In such cases, if enough mimics successfully signaled that they were poisonous, it could be advantageous for a predator to ignore the signal and feed on the excess mimics. Assigning classes in this way also helped stabilize the system, preventing the extinction of any one prey class.

Because predators could not act as a top-down control on the poisonous prey class, we limited the number of prey in each class to 1,000 organisms and imposed the following method of class-specific population size limitations. Whenever an organism was born, it was assigned to a prey class. If the inclusion of the newly born organism would increase the number of prey in that class beyond the prey type cap, a random existing individual from that class was removed from the population. This method follows the same logic used in Avida by default in which population sizes are limited by physical space constraints, except that we applied independent limits to each class, instead of to the population as a whole. We set resource inflow levels sufficiently high to ensure that they did not directly limit prey population levels. Consequently, cases in which a prey class population was substantially less than 1,000 specifically indicated that the class was top-down limited by predators. Unlike the three prey classes, predators were not limited to a maximum class size: predator resources (i.e. prey) were always finite and in a negative frequency relationship with the predator population size.

All populations were evolved for 500,000 updates (an update is an arbitrary time unit in Avida, roughly equivalent to the time required for each organism in the population to execute 30 genomic instructions). We utilized identical configuration parameters for all treatments, only varying the efficacy of the poison by manipulating the poison level associated with the poison prey (levels used: 0.01, 0.025, 0.05, 0.1, 0.2, 0.4, 0.75, and 1.0), as specified in the results. We conducted 30 replicate runs for each treatment. In practice, early population extinctions (due to excessive predation) reduced sample sizes in four treatments: at the 0.01, 0.025, 0.05, and 0.1 poison efficacy levels, 8, 2, 2, and 2 of the populations were lost and excluded from our analyses.

We used Avida version 2.12 for all experiments. Data were post-processed using Python 2.7.1 and the pylab and matplotlib libraries from the Enthought Python Distribution version 7.02-2. Statistical analyses and plotting were conducted in R version 2.15.2 using the libraries ggplot2 version 0.9.3.1, gridExtra version 0.9.1, and boot version 1.3-5.

## Results

We tested for levels of toxicity necessary for the evolutionary emergence of signal recognition from a cue by predators and dishonest signaling by prey mimics. We also tested the level of mimic accuracy required to support a successful mimic population.

### *Predator Recognition of a Cue*

We assessed the conditions under which predators evolved to preferentially avoid consuming poisonous prey using a logistic regression model that related predation levels to the eight levels of toxicity of the poisonous prey (0.01, 0.025, 0.05, 0.1, 0.2, 0.4, 0.75, 1.0). In this model, we classified each of the 30 poison prey sub-populations as being under predation pressure if their realized abundance fell below 800 individuals (80% of the maximum) at the end of the experimental trial. This threshold was chosen arbitrarily during preliminary data analysis in an effort to reduce noise as experimental abundances were either significantly above or significantly below this value, but the individual abundance values were highly stochastic.  Based on the proportion of evolved populations in which poison prey were not under predation pressure, the model predicts the probability that predators will evolve to preferentially avoid consuming poisonous prey at a given a poison efficacy level (Figure 1). At poison levels below 0.1 the selective pressure to avoid such prey is weak, resulting in a low probability that predators evolved selective predation habits. However, when the efficacy of the poisonous prey is at or above 0.1, predator populations nearly universally evolve to avoid predating poisonous prey. To illustrate the potential realized cost of consuming a poisonous prey, a poison level of 0.1 would cause a 10% reduction in the resources available for an attacking predator to satisfy the threshold for reproduction (10 units). As such, in order to compensate for 10% 'energetic' loss, if a predator had previously stored 9 units, it would now have to consume two young prey that each had consumed 1 unit of environmental resource, or one better-fed prey that had consumed 2 units. On the other hand, had that same predator killed a non-poisonous species, it could reproduce immediately.

*Selective Targeting of Prey Types*

For all tested poison levels, the three prey classes rapidly grew toward the population cap until predators began exerting top-down controlling pressure (Figure 2). At low poison levels (Figures 2a and 2b), mimic and safe prey subpopulation sizes remained relatively comparable and constant throughout evolution, suggesting that predators were consuming prey in proportion to their availability and not distinguishing between poisonous and non-poisonous phenotypes. However, at higher poison levels (Figures 2c and 2d), the poisonous prey subpopulations converged to the maximum sub-population size, indicating that they were no longer under top-down predation control and that predators had evolved selective targeting of non-poisonous prey. At the same time, realized predator population sizes were higher in high poison efficacy trials. This indicates that predators could have benefited from evolving skills for discriminating prey at low poison levels. However, at these levels, the selection pressures were not (apparently) strong enough for predator populations to realize this potential in the allotted evolutionary time.

*Evolved Manipulation of a Communicative Signal*

In all populations, prey evolved to alter their apparent phenotypic signal (i.e. appearance) when in the mimic class (Figure 3). However, at low poison levels (0.025 and 0.05, Figure 3a and 3b, respectively), mimic class prey did not demonstrate a clear preference for mimicking any particular prey type. In contrast, at high poison levels (0.1 and 0.2, Figure 3c and 3d, respectively), mimic prey showed a clear preference toward mimicking poisonous prey. Thus it was only when poison efficacy was high and predators evolved to selectively avoid poisonous prey (Figure 1) that prey evolved to selectively mimic that prey type.

*Signal Noise and the Effect of Information Loss*

In the experiments considered above, the phenotypic appearance of mimics was perceived with perfect fidelity by the predators. I.e. predators always saw what the mimics intended them to see. In order to test the robustness of strategies for mimicking poisonous prey, we further evaluated mimicry choices by evolving populations under 'imperfect' mimicry conditions. Under imperfect mimicry, predators perceived the intended mimic signal only 25% of the time, with the true (mimic) phenotype apparent to the predator the rest of the time. From the final populations, we then calculated the ratio of the mean proportion of organisms of each population that were mimicking poisonous prey under the low accuracy mimicry conditions to the mean proportion mimicking poisonous prey when mimicry was perfect for each of the five poison levels (0.10, 0.20, 0.40, 0.75, 1.0; Figure 4). By this measure, ratios over 1.0 would indicate a greater proportion of organisms were attempting to mimic poisonous prey when mimicry was imperfect than when it was perfect. We calculated 95% bootstrap confidence intervals for these measures by repeatedly calculating the ratio from sampled subsets of the source populations (i.e. 25% and 100% accuracy populations) over 1000 iterations. At the two highest poison levels, 0.75 and 1.0, the ratios (95% CIs) were 1.28 (1.20 – 1.37) and 1.39 (1.29 – 1.47), respectively. The three lower poison levels, 0.1, 0.2, and 0.4, had realized ratios of 0.71 (0.66 - 0.76), 0.83 (0.77 – 0.89), 0.81 (0.75 – 0.86), respectively. Overall, these data indicate that low proportions of organisms in the mimic class chose to mimic poisonous prey when mimicry was imperfect (i.e. most ratios fell below 1.0). However, high poison efficacy (>= 0.75) provided enough protection to poison prey phenotypes that a higher proportion of prey in the mimicry class had evolved to appear poisonous, even though mimicry was highly imperfect. Values above 1.0 indicate that at high poison levels a lower proportion of perfect mimics mimicked the poison model than did so under imperfect mimicry conditions. We suggest that may have arisen because, under perfect mimicry, predators should specialize only on the non-poisonous prey class due to the high cost of misidentifying the poisonous model as a mimic. This reduces selection pressures for mimics to mimic the poisonous phenotype. However, when mimicry is imperfect, the edible population of mimic prey remains more numerous and may therefore sustain predators specializing on the mimic class. Under such conditions, selection will more strongly favor mimics that avoid these specialists by mimicking the poison model.

## Discussion

We have demonstrated that adequate toxicity is required for aposematic cue recognition to evolve and inform predatory behavior. At low poison levels, predators distinguished between prey types in only 3 of 26 trials. However, at higher poison efficacy levels, predator recognition and selection of prey types increased, with the behavior fixing in all trial populations for poison levels at and above 0.4. This agrees with previous findings that predator learning is enhanced by highly distasteful prey [21]. At the same time, selective pressures on prey were strong enough to promote the evolution of dishonest signaling through mimicry of the aposematic signal (Figure 3). Mimics and dishonest signaling did not cause predators to ignore the aposematic cue (Figures 2). Instead, at the given prey type 'immigration' rates used here, mimics evolved to deceptively signal that they were poisonous without significantly destabilizing predator cue recognition (Figure 3). Additionally, once predators began to cue in on and respond to prey signals, at higher poison-prey toxicity levels, higher proportions of mimics signaling that they were poisonous, leading to an increase in mimic survival relative to safe prey (Figure 2).

Despite being capable of mimicking poisonous species, mimics did not universally evolve to employ this strategy. Instead, as has often been observed in nature and is predicted in mathematical models [20,55,56], the ratio of expressed mimic signals appears to reach equilibrium. This occurred at all poison levels, suggesting that the toxicity of the model can only provide protection to a certain number of mimics and that this level of protection is governed by negative frequency dependence. Any non-poisonous species that becomes highly abundant will experience increased selection pressure, eventually driving it into rarity while other prey species populations increase, causing an increase in predation on that species and ultimately leading to stable signaling ratios, as are apparent in Figure 3.

Brower's model [25] demonstrated that highly toxic prey can support an abundance of perfect mimics. Similarly, we found that the proportion of mimics signaling toxicity increased as the toxicity of the model increased. Under conditions of imperfect mimicry (Figure 4), predators increasingly generalize the signals of toxic prey as the prey's toxicity increases, because the cost of failing to identify a model as a mimic is too high [39,57], particularly in the presence of alternative prey [45]. However, our experiments also show that in environments with coevolving naïve predators, imperfect mimicry is supported without requiring high levels of toxicity. This finding lends support to Fisher's theory [4] of gradual evolution of mimicry. Our results suggest that other dishonest signals may have evolved gradually in situations where the cost of incorrectly distinguishing a dishonest from honest signal is high. Similarly, in the coral snake mimicry complex, the most perfect mimics appear at the edge of the model range [58] and high model abundance at the center supports imprecise mimicry because of predatory generalization [39]. At the same time, our findings of predator generalization and the evolution of imperfect mimicry contrast with the assertion that mimicry must evolve in a two-step process that starts with feature saltation [37,59]. Such feature saltation would allow a species to jump the adaptive valley between crypsis and mimicry, and then gradually evolve toward the adaptive peak defined by the model's appearance [33,59]. Our results suggest that this two-step process is not a necessary mechanism for the evolution of mimicry systems.

Overall, we have demonstrated that mechanisms of communication based on cue recognition readily evolve when it provides adequate benefits to both parties. Further, we have shown that these systems are robust to high levels of noise: cues, once recognized, can support the evolution of dishonest mimicry signaling, even when the mimicry signal is highly imperfect, without disrupting the communication system. Understanding how basic signaling systems evolve can help us understand the selective pressures leading to more complex communication and language systems. Our findings suggest that communicative signaling systems can evolve readily and gradually, without feature saltation, and can confer adaptive advantages allowing populations to cross adaptive valleys toward increasingly sophisticated signal-receiver communications systems.

**Acknowledgments**

We are grateful for the invaluable input of Chris Adami, Mike Wiesenauer, and the BEACON 2013 class. This work was supported by the BEACON Center for the Study of Evolution in Action (NSF Cooperative Agreement DBI-0939454) and the Michigan State University Institute for Cyber Enabled Research. Any opinions, findings, and conclusions or recommendations expressed in this material are those of the author(s) and do not necessarily reflect the views of the National Science Foundation.

**Figure 1. Predators evolve to recognize and avoid consuming poisonous prey, even**

**at relatively low poison efficacy levels.** Data shown represent a fit from a logistic regression model relating poison level to the probability that predator species will evolve to avoid consuming poisonous prey (based on proportion of evolved populations in which poison prey were no longer under predation pressure). Solid black line indicates the predicted probability. The red dashed lines represent the 95% bootstrap confidence intervals of the model. Circles indicate the observed values in our experiments.

**Figure 2. Predators benefit from evolving to distinguish and preferentially avoid poisonous prey at high poison efficacy levels.** Shown are mean population size for the three prey classes, non-poisonous (blue), poisonous (pink), and mimics (orange), as well as predator abundance (grey) at the four tested poison levels, 0.025 (a), 0.05 (b), 0.1 (c), and 0.2 (d). Shaded regions show the 95% bootstrap confidence intervals calculated from 10,000 iterations.

**Figure 3. Prey preferentially mimic poisonous prey under conditions of high poison efficacy and associated evolved predator selectivity.** Shown are proportion of mimic class organisms presenting phenotypes representing each available prey class: mimic (orange), non-poisonous (blue), and poisonous (pink) at the four tested poison levels, 0.025 (a), 0.05 (b), 0.1 (c), and 0.2 (d). Shaded regions show the 95% bootstrap confidence intervals calculated from 10,000 iterations.

**Figure 4. High poison efficacy levels promote the evolution of mimicry, even when mimicry is highly imperfect.** Shown are mean ratios of the mean proportion of organisms in each population that were mimicking poisonous prey under low accuracy mimicry conditions (25%) to the mean proportion mimicking poisonous prey when mimicry was perfect (100% accuracy). The observed ratios greater than 1.0 indicate that selection to mimic poisonous prey was even greater at higher poison levels when mimicry was imperfect than it was when mimicry was perfect because increased toxicity supports imperfect mimicry. At low poison levels, a lower proportion of organisms in the mimic class mimic poison prey when mimicry is imperfect. CI's given are 95% bootstrapped confidence intervals. Bootstrapping was performed by repeatedly calculating the ratio from sampled subsets of the source populations (i.e. 25% and 100% accuracy populations) at 500,000 updates.

| Instruction Sequence | Non-poisonous | Mimic | Poisonous |
|---|---|---|---|
| attack | ● | ● | ● |
| attack + nop-A | ● | | |
| attack + nop-B | | ● | |
| attack + nop-C | | | ● |
| attack + nop-D | ● | | |
| attack + nop-E | | ● | |
| attack + nop-F | | | ● |
| attack + nop-G | ● | | |
| attack + nop-H | | ● | |

**Table 1. Attack instruction sequence targeting.** The prey type targeted by each of nine possible attack instruction sequences are shown above, indicated by the ● symbol.

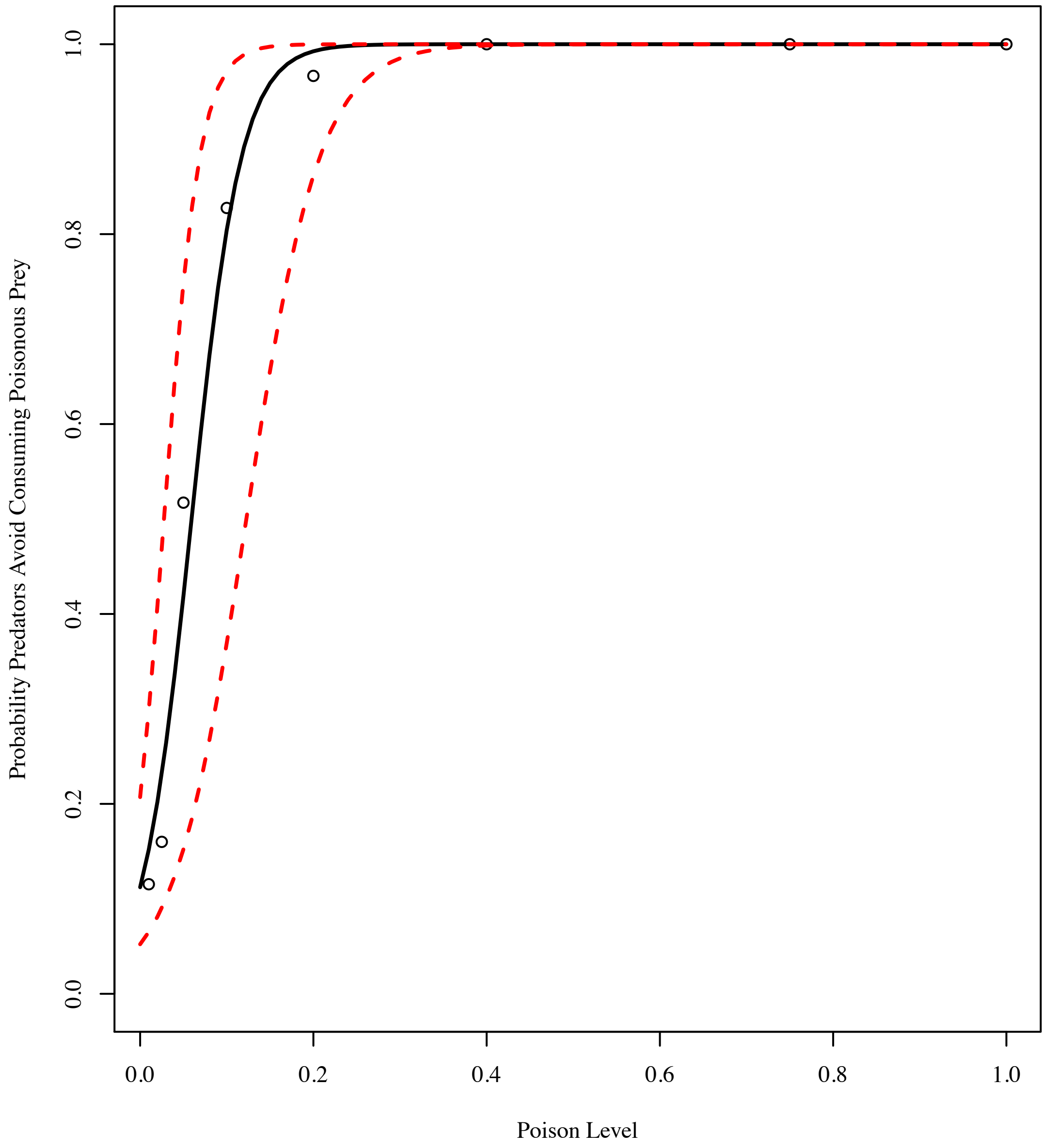

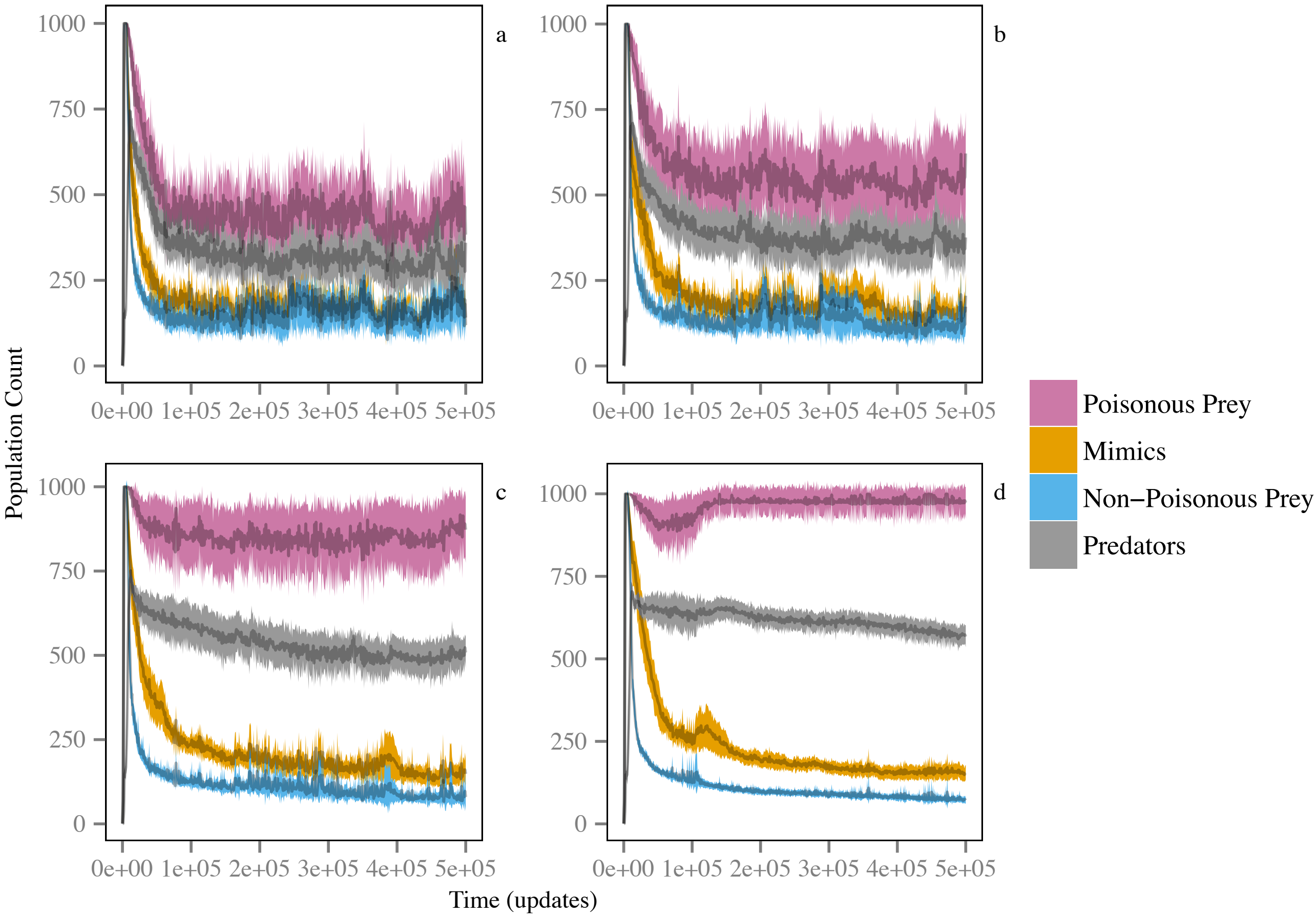

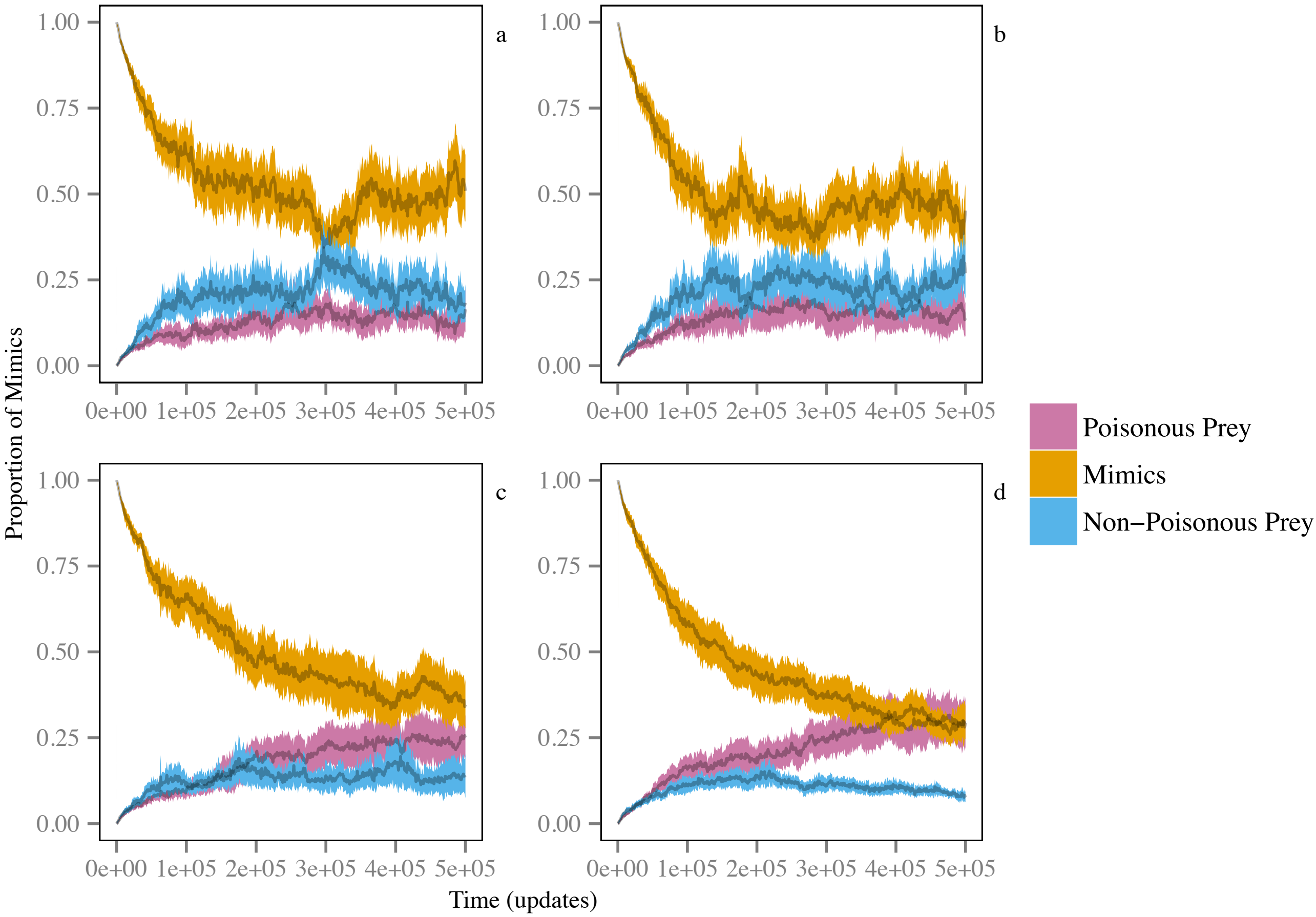

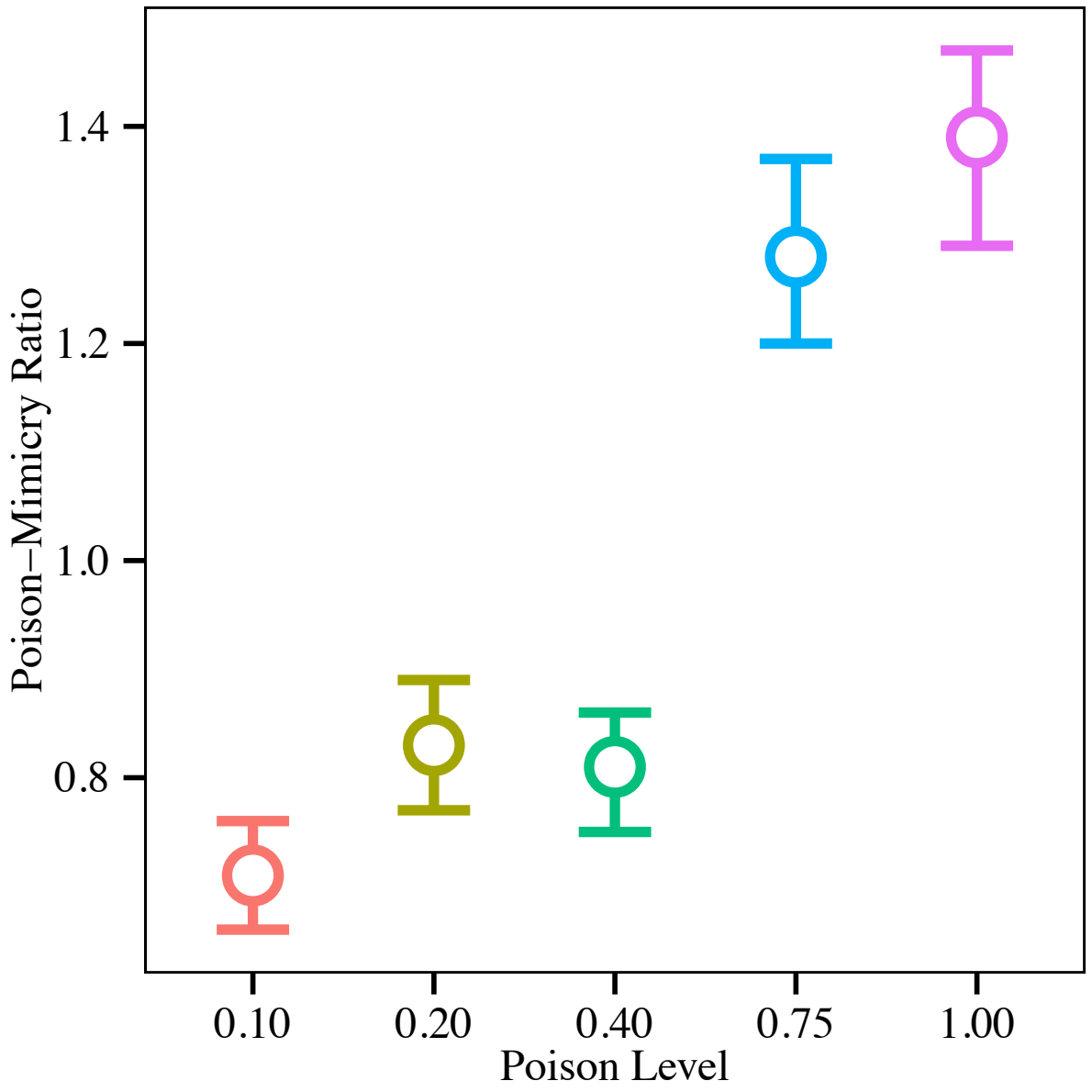